\documentclass[reprint,aps,prb,superscriptaddress,a4paper,nofootinbib,showpacs,showkeys,floatfix]{revtex4-1}
\usepackage{url}
\usepackage{graphicx,epsfig,epstopdf}
\usepackage{amssymb,amsmath,bm,dcolumn,braket,url,color}
\usepackage[normalem]{ulem}

\newcommand{\ea}{{\it et al.}}

\begin{document}

\pagenumbering{arabic}




\bibliographystyle{apsrev4-1}

\title{Local geometry around B~atoms in B/Si(111) from polarized 
  x-ray absorption spectroscopy}

\author{Saleem Ayaz \surname{Khan}} \affiliation{New Technologies
  Research Centre, University of West Bohemia, Univerzitn\'{\i} 2732,
  306 14 Pilsen, Czech Republic}

\author{Martin \surname{Vondr\'{a}\v{c}ek}} \affiliation{Institute of
  Physics ASCR v.~v.~i., Na Slovance 2, CZ-182~21 Prague, Czech Republic}

\author{Peter \surname{Blaha}} \affiliation{Institute of Materials
  Chemistry, TU Vienna, Getreidemarkt 9, A-1060 Vienna, Austria}

\author{Kate\v{r}ina \surname{Hor\'{a}kov\'{a}}} \affiliation{Institute of
  Physics ASCR v.~v.~i., Na Slovance 2, CZ-182~21 Prague, Czech Republic}

\author{Jan \surname{Min\'{a}r}} \affiliation{New Technologies
  Research Centre, University of West Bohemia, Univerzitn\'{\i} 2732,
  306 14 Pilsen, Czech Republic} 

\author{Ond\v{r}ej \surname{\v{S}ipr}} \affiliation{Institute of Physics
  ASCR v.~v.~i., Cukrovarnick\'{a}~10, CZ-162~53~Prague, Czech
  Republic }

\author{Vladim\'{\i}r  \surname{Ch\'{a}b}} \affiliation{Institute of
  Physics ASCR v.~v.~i., Na Slovance 2, CZ-182~21 Prague, Czech
  Republic}



\date{\today}

\begin{abstract}
The arrangement of B~atoms in a doped
Si(111)-$(\sqrt{3}\times\sqrt{3})R30^{\circ}$:B system was studied
using near-edge x-ray absorption fine structure (NEXAFS). Boron atoms
were deposited via segregation from the bulk by flashing the sample
repeatedly. The positions of B~atoms are determined by comparing
measured polarized (angle-dependent) NEXAFS spectra with spectra
calculated for various structural models based on ab-initio total
energy calculations.  It is found that most of boron atoms are located
in sub-surface L$_{1}^{c}$ positions, beneath a Si atom.  However, depending on
the preparation method a significant portion of B~atoms may be located
elsewhere.  A possible location of these non-L$_{1}^{c}$-atoms is at the
surface, next to those Si atoms which form the
$(\sqrt{3}\times\sqrt{3})R30^{\circ}$ reconstruction.
\end{abstract}


\maketitle


\section{Introduction}

Studies of the B/Si(111) system started intensively in the late
eighties \cite{Lyo+89,Headrick+89}. After a period of a declined
interest, the research has again intensified stimulated by attempts to
prepare a passivated Si surface in connection with the development of
molecular electronics. Full introduction of molecules into the
technology is still in its initial stages, related to a production of
hybrid circuits composed of parts produced with Si-based and organic
technologies. One of the challenges in this field is tailoring the
interaction between deposited molecules and a substrate that is needed
for wiring in a device. Employment of the B/Si(111) system is very
promising as it can be prepared with different technologies
(segregation or epitaxy) and with different properties: either as a
spacer or as a passivated surface layer in the form of
$\delta$-doping. The latter one offers a surface with active isolated
Si atoms that can be considered as centres for molecule capturing.
This view is supported, e.g., by a recent theoretical work
concentrating on the interaction of various metalphthalocyanine (MPc)
molecules with the $\delta$-doped
Si(111)-$(\sqrt{3}\times\sqrt{3})R30^{\circ}$:B surface: for some
molecules this interaction has van der Waals character that
enables diffusion of the molecules on the surface so that
self-organized structures can be formed \cite{Veiga+93}.

The location of B~atoms at the Si(111) surface was carefully examined
in the past.  Low-energy electron diffraction (LEED) and similar
methods showed the common $(\sqrt{3}\times\sqrt{3})R30^{\circ}$
surface where boron atoms might be located in the T or L$_{1}^{c}$ positions
\cite{Akimoto+90,Huang+90,Baumg+99}, well defined in the
dimer-adatom-stacking fault model of the Si(111)-$7\times7$
surface.  The most accepted
position is the L$_{1}^{c}$ site, with Si atoms on top of B~atoms in the second
layer. This conclusion has been supported by calculations of total
energies for different structural models
\cite{Lyo+89,BMM+89,KPH+90,Chang+97,Shi+02,Andrade+15}.

Despite the results obtained so far, the question where the B~atoms
are located cannot be regarded as settled.  The intensity of
diffraction spots represents data averaged over different
configurations that cannot be identified in detail.  It is conceivable
that local configurations that cannot be distinguished by the
diffraction are present.  Indeed, several local-probe studies
involving scanning tunneling microscopy (STM) or atomic-force
microscopy (AFM) suggest that the B~atoms may occupy also other
positions than the L$_{1}^{c}$ site
\cite{Shen+94,Zotov+96,stimpel,Spadafora+14}.  A lot of attention was
focused on how the structure varies depending on the conditions of
preparation, especially on the heat treatment
\cite{Lyo+89,Shen+94,Zotov+96,stimpel,NMS+00,KOF+11}.

Recently a combined experimental (STM) and theoretical study showed
that there can be two charge states and consequently two local L$_{1}^{c}$
configurations for the $\delta$-doped Si(111) surface owing to
electron--lattice coupling \cite{EOM}.  One state corresponds to the
ground state of the $(\sqrt{3}\times\sqrt{3})R30^{\circ}$
reconstruction while the second state is a two-electron bound state
with an elevated Si adatom. The possibility of switching between these
states has been found at low temperatures ($T<70$~K). Note that the
concept of two concurrent dynamically switchable geometries has been
extensively employed in modelling the Si(100)-$2\times1$ reconstructed
surface \cite{TFD+92,RBB+99,SJD+11}.

To learn more about the positions of B~atoms at Si(111) it is
desirable to employ a local method which, unlike the STM, probes a
part of the sample large enough to be considered as truly
representative.  The x-ray absorption spectroscopy satisfies these
needs: it is chemically specific, meaning that one can be sure that it
is the nearest neighborhood of a B~atom which is considered, and at
the same time the area inspected is macroscopic (typically
0.1~$\text{mm}\times0.5$~$\text{mm}$).  As concerns the
theoretical approach, a potentially weak point of all previous studies
is that they employed pseudopotentials. Such calculations are
computationally efficient but a verification of the results by an
all-electron method is always desirable.

In this study we present experimental near-edge x-ray absorption fine
structure (NEXAFS) spectra measured at the B $K$-edge for boron
$\delta$-doped Si(111)-$(\sqrt{3}\times\sqrt{3})R30^{\circ}$ surface,
prepared by flashing at two different temperatures (1100~$^{\circ}$C
and 900~$^{\circ}$C).  The data are simulated using the all-electron
{\sc wien2k} code considering several trial geometries suggested by
total energy minimization.  By comparing experimental and theoretical
NEXAFS spectra we found that the B~atoms are mostly in the L$_{1}^{c}$
positions. Depending on the preparation method, however, a significant
portion of the B~atoms may be in different positions, possibly in the
surface L$_{1}^{a}$ site, next to those Si atop atoms which form the
$(\sqrt{3}\times\sqrt{3})R30^{\circ}$ reconstruction.

\section{Methodological framework}

\subsection{Experiment}

\begin{figure}
\begin{tabular}{@{}cc@{}}
\includegraphics[width=0.50\linewidth]{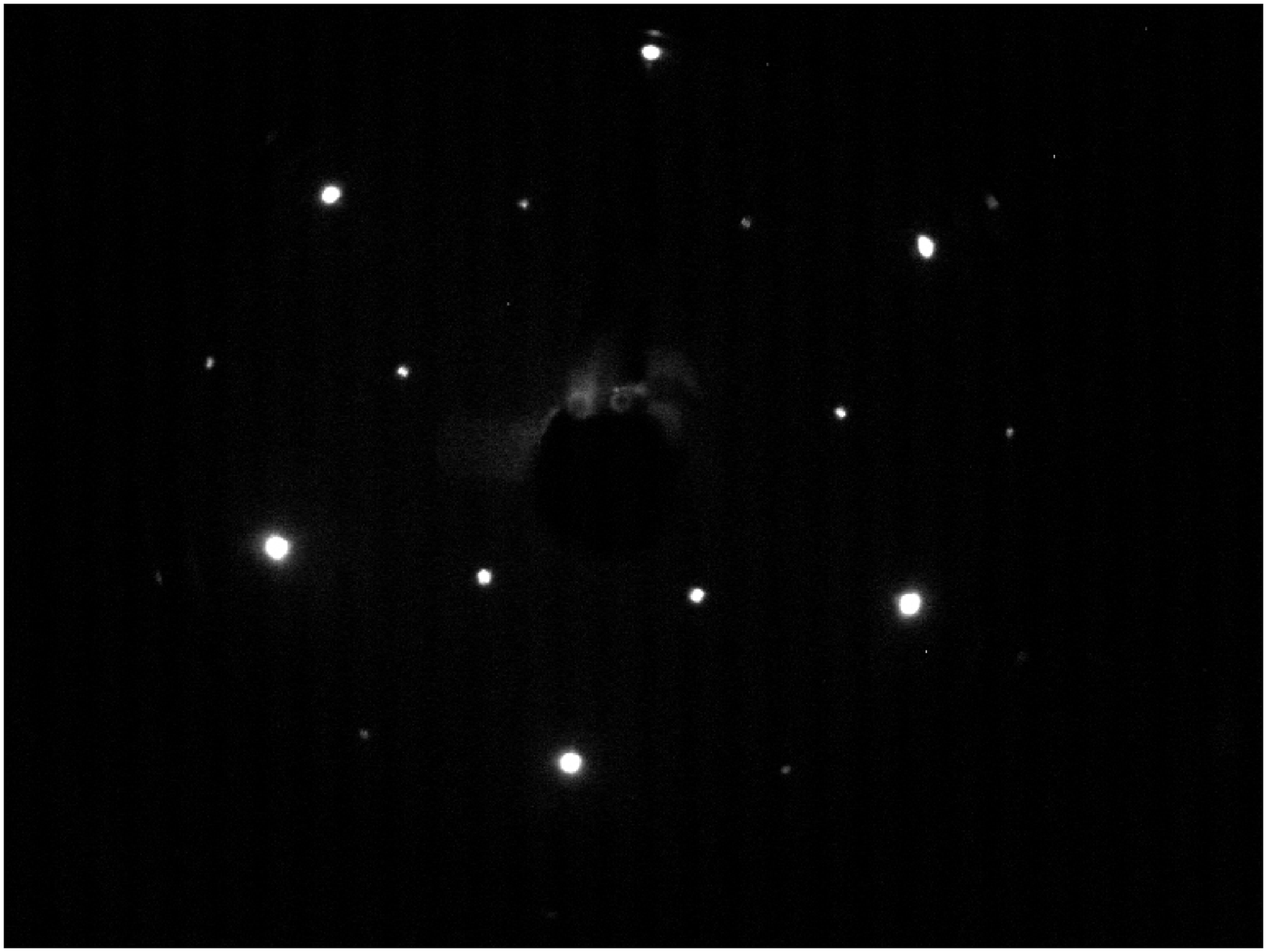} &
\includegraphics[width=0.59\linewidth]{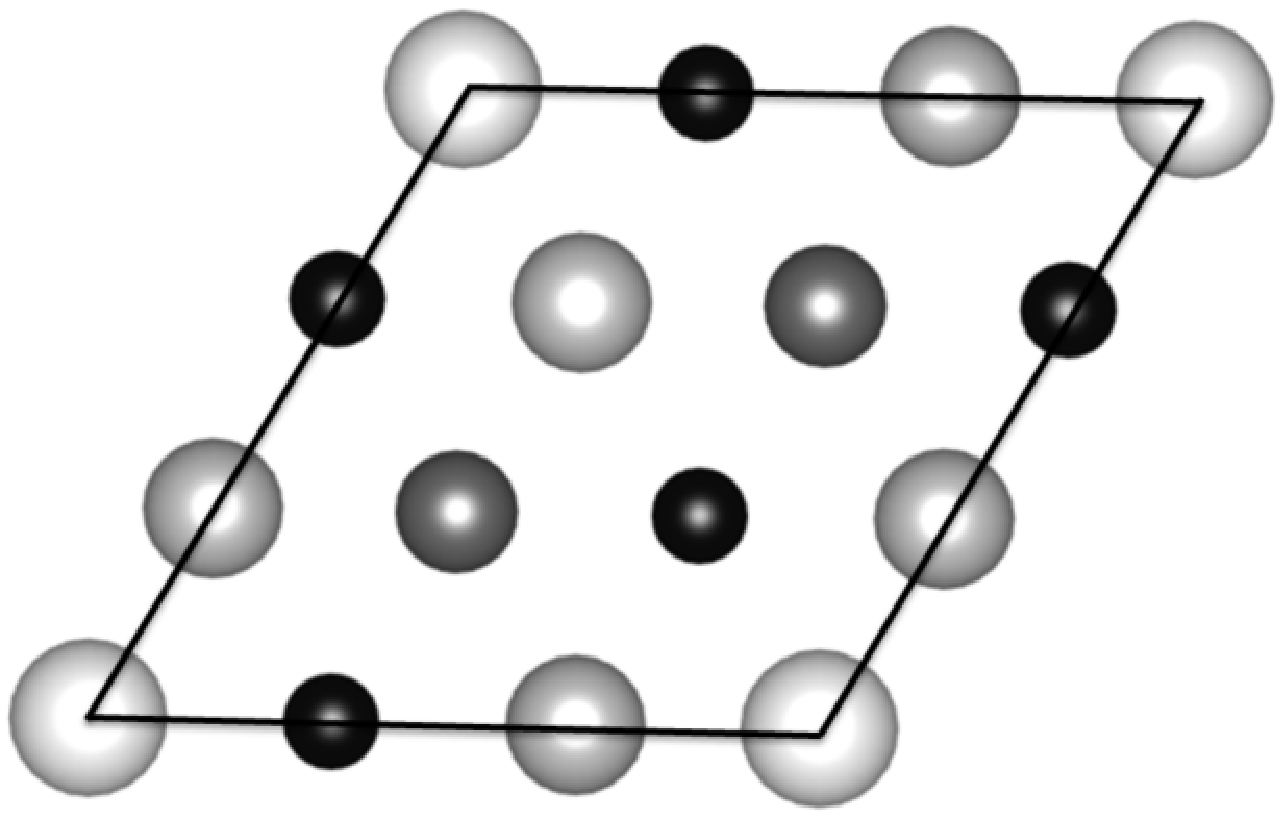}
\end{tabular}
\caption{ (a) LEED of surface reconstruction
  $(\sqrt{3}\times\sqrt{3})R30^{\circ}$ for the B/Si(111) samples. (b) Schematic top view of the Si(111)
  $(\sqrt{3}\times\sqrt{3})R30^{\circ}$ reconstructed
  surface.  Different size of atoms in different layers are used
  to provide a better insight.
 \label{fig1} }
\end{figure}

The samples were prepared by segregation of B atoms from the bulk:
First, a clean Si surface was obtained by annealing a highly B-doped
Si(111) wafer (resistivity less then 0.01~$\Omega$cm,
$N_{A}\sim10^{19}$~cm$^{-3}$) for 12~hours in ultra high vacuum at the
temperature 500~$^{\circ}$C and pressure $8\times10^{-10}$~mbar.  To
achieve B atoms segregation, the samples were repeatedly flashed for
5~s at temperatures 1100~$^{\circ}$C (denoted as sample 1100) or
900~$^{\circ}$C (denoted as sample 900); the pressure was maintained
less then $8\times10^{-9}$~mbar.  After this procedure, a surface
reconstruction $(\sqrt{3}\times\sqrt{3})R30^{\circ}$ has been
identified by LEED, as seen on Fig.~\ref{fig1}(a).

B $K$-edge NEXAFS spectra were recorded at the Materials Science
Beamline, Elettra Sincrotrone Trieste, Italy \cite{Vasina+01}.  The
data were acquired via surface sensitive Auger electron yield
measurements, by recording the intensity of the B~$KLL$ Auger
transition. The angle between the photon beam and the axis of the
electron analyzer SPECS Phoibos 150 was fixed to 60$^{\circ}$ and the
sample was rotated around the vertical axis. The NEXAFS spectra were
acquired at four angles, ranging from normal incidence where the
polarization vector $\varepsilon$ is in the Si(111) (or $xy$) plane
through the 30$^{\circ}$ incidence angle and normal emission angle
(60$^{\circ}$ incidence) to the grazing incidence at 80$^{\circ}$,
with $\varepsilon$ nearly parallel to the surface normal.  The overall
energy resolution for measured B $K$-edge NEXAFS spectra was 0.2~eV.


\subsection{Structural models}
\label{sec-models}

\begin{figure}
\begin{center}
\includegraphics[width=0.7\linewidth]{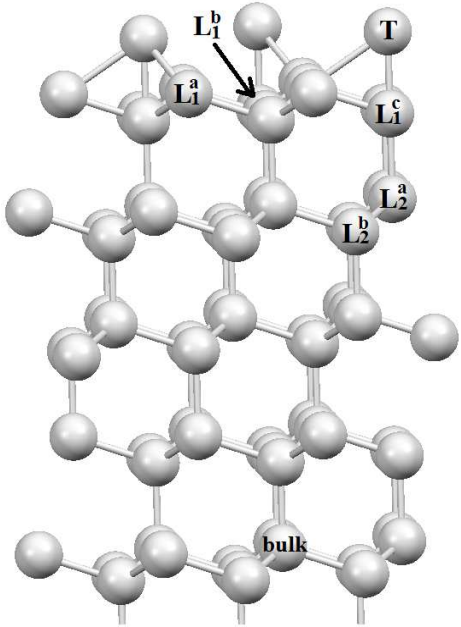}
\caption{Different configurations of the B~atom at Si(111)
  $(\sqrt{3}\times\sqrt{3})R30^{\circ}$ \label{fig2} }
\end{center}
\end{figure}

The system is modeled by a supercell of slabs. Each slab
  consists of seven layers of Si atoms, with an additional incomplete
  layer of topmost Si atoms.  Hydrogen atoms
  were added to saturate the dangling bonds at the other side of the
  slab. The thickness of the slab is about 23~\AA.  In the supercell
  the slabs are separated by about 14~\AA\ of vacuum.  Concerning the
  horizontal geometry, the slabs were constructed so that they
  correspond to the $(\sqrt{3}\times\sqrt{3})R30^{\circ}$
  reconstruction which the B/Si(111) system undergoes (see the diagram
  in Figure~\ref{fig1}(b)).

The structural models we explored were chosen by considering several
positions of B~atoms based on the ab-initio structural study of Andrade
\ea\ \cite{Andrade+15}.  These positions are depicted schematically in
Fig.~\ref{fig2}, where we show five upper layers and the
$(\sqrt{3}\times\sqrt{3})R30^{\circ}$ reconstruction atom at the top.
We adopted a nomenclature that highlights the location of sites
in specific layers, starting from the top ($T$, $L_{1}$, $L_{2}$).  Our
study considers not only the L$_{1}^{c}$ position (labeled as S5 by
Andrade \ea\ \cite{Andrade+15}) which attracted most attention in
earlier works but also several other positions, which energetically
least deviate from the L$_{1}^{c}$ geometry and/or which should be
considered based on kinematic reasons.

The structure relaxation was performed so that first the structure of
bulk Si crystal was optimized to obtain the optimized bulk Si-Si distance
(2.397~\AA).  This interatomic distance was then set as fixed for atoms in
the two lowermost layers of our slab.  The positions of the other
atoms were optimized by allowing the atoms to move in the direction of
the force untill the equilibrium has been attained.


\subsection{Calculations}

The spectra were calculated by the ab-initio all-electron full
potential linear augmented plane wave (FLAPW) method, as implemented
in the {\sc wien2k} code \cite{Blaha+01}. 
 The calculations were performed using the Perdew, Burke and
  Ernzerhof generalized gradient approximation (PBE-GGA)
  exchange-correlation functional \cite{Perdew+96}.  Additionally, we
  employed also the meta-GGA SCAN functional \cite{Sun+15} to evaluate
  the total energies for structures that have been already optimized
  via the PBE functional.  This step is motivated by the fact that the
  SCAN functional often improves the energetics (while the atomic positions
  are usually well-predicted already with the PBE functional)
  \cite{TSB+16}.

Wave functions in the interstitial regions were expanded in plane waves,
with the plane wave cutoff chosen so that $R_{\text{MT}}K_{max}$=5 (where
$R_{\text{MT}}$ represents the smallest atomic sphere radius and $K_{max}$ is the
magnitude of the largest wave vector). The $R_{\text{MT}}$ radii were
taken as 1.78~a.u.\ for Si atoms, 1.80~a.u.\ for B~atoms and
0.95~a.u.\ for H atoms. The wave-functions inside the spheres were
expanded in spherical harmonics up to the maximum angular momentum
$\ell_{\text{max}}$=10. The $\bm{k}$-space integration was performed
via a modified tetrahedron integration scheme.  The internal geometry
of the system is optimized using 2~$\bm{k}$-points in the irreducible
Brillouin zone (IBZ) distributed according to a $(2\times2\times1)$
Monkhorst-Pack grid\cite{Monkhorst+76} while the self consistencies of
the ground state energies were obtained by 8~$\bm{k}$-points in IBZ,
distributed according to a $(4\times4\times1)$ Monkhorst-Pack grid.

Polarized x-ray absorption spectra were calculated via Fermi's
Golden rule within the dipole approximation \cite{Mandl+92}.  The raw
spectra were convoluted by a Gaussian with full width at half maximum
(FWHM) of 0.3~eV and by a Lorentzian with FWHM of 0.2 eV, to simulate
the effect of the experimental broadening and of the finite core hole
lifetime.  The differences between the NEXAFS for the $\varepsilon\|x$
and $\varepsilon\|y$ polarizations are very small, therefore we always 
display just their average. We distinguish in the following only
between spectra with polarization vector in-plane (normal incidence)
or out-of-plane (grazing incidence).  There is a small difference
between what is considered as grazing incidence in experiment and
theory.  In the experiments the grazing incidence means that the
incoming radiation arrives at the sample not truly parallel to the
surface but at an angle of 10$^{\circ}$; the polarization vector is
thus tilted by 10$^{\circ}$ from the normal.  In the calculations we take 
the polarization vector exactly parallel to the surface normal.  We do
not expect any significant differences between spectra for the
``true'' and ``approximative'' grazing incidence setups.

The influence of the core hole on B $K$-edge NEXAFS can be
considerable \cite{SR+10}.  It can be accounted for via one of the
approximative static schemes.  Frequently one relies on the final
state rule \cite{Barth+82}, meaning that the spectrum is evaluated for
electron states which have relaxed to the presence of the core hole.
To employ this scheme, we performed first a self-consistent
calculation with one 1$s$ electron removed from the B~atom and at the
same time with one electron added to the valence states to maintain
the charge neutrality.  After the self-consistency had been achieved,
another ``single-shot'' calculation was performed with the additional
electron removed from the valence states, to get a proper Fermi level.
We did not introduce another (larger) supercell scheme in this
respect, because the B~atoms are already quasi-isolated for the
reconstructed $(\sqrt{3}\times\sqrt{3})R30^{\circ}$ system --- their
distance is 6.7~\AA.

It is difficult to guess {\em a priori} which way of dealing with the
core hole is the most suitable for a particular situation, therefore,
we performed exploratory calculations for several core hole schemes:
we calculated the NEXAFS (i) using a ground state potential (no core
hole), (ii) using a potential obtained via the final state rule as
described above and (iii) using a potential obtained via a final state
rule with half of a core hole, which is equivalent to relying on
Slater's transition state approximation.  Following the outcome for one
particular geometry (see appendix~\ref{sec-corehole}), we decided to
use the final state rule approximation with a full core hole
throughout this study.


\section{Results}

\label{sec-results}


\subsection{Comparing total energies}

\label{sec-ene}

\begin{table}
\caption{Total energies for systems with B~atoms in positions depicted
  in Fig.~\protect\ref{fig2}.  The values are given relative to the energy of
  the system with the B~atom in the L$_{1}^{c}$ position. Total
  energies of Andrade \ea\ \cite{Andrade+15} are shown for comparison
  (together with their notation for the positions of the B
  atoms). \label{tab-totene}}
\begin{ruledtabular}
\begin{tabular}{ccccc}
  \multicolumn{2}{c}{position of B atom}
  & \multicolumn{3}{c}{$\Delta E$\ (eV)} \\
  notation & \multicolumn{1}{c}{notation} & \multicolumn{1}{c}{PBE} &
  \multicolumn{1}{c}{SCAN} & \multicolumn{1}{c}{PBE} \\ 
  present & Andrade & \multicolumn{1}{c}{present} &
  \multicolumn{1}{c}{present} & \multicolumn{1}{c}{Andrade} \\ 
\hline 
L$_{1}^{c}$  & S5     &  0.00  &  0.00  &   0.00  \\ 
L$_{2}^{a}$  & B$_{1}$ &  0.55 &   0.68  &  0.39  \\ 
L$_{2}^{b}$  & C      &  0.91  &  1.11  &   0.76  \\ 
L$_{1}^{a}$  & T$_{5}$ &  1.14 &   1.31  &  1.05  \\ 
L$_{1}^{b}$  & A       & 1.18  &  1.25  &   1.21  \\ 
T           & T$_{4}$ &  1.28  &  1.46  &  1.22  \\ 
bulk        & --     &  1.41  &  1.75  &   --    \\
\end{tabular}
\end{ruledtabular}
\end {table}


Total energies obtained for B~atoms in positions depicted in
Fig.~\ref{fig2} are presented in Tab.~\ref{tab-totene}.  For
comparison, we show also the results of earlier pseudopotential
calculations of Andrade \ea \cite{Andrade+15}.
It follows from Tab.~\ref{tab-totene} that using the
  meta-GGA SCAN functional leads to the same trends as obtained for
  the GGA PBE functional --- except for the  L$_{1}^{a}$ and  L$_{1}^{b}$
  positions where the trend is reversed. 

\begin{figure}
\includegraphics[width=0.95\linewidth]{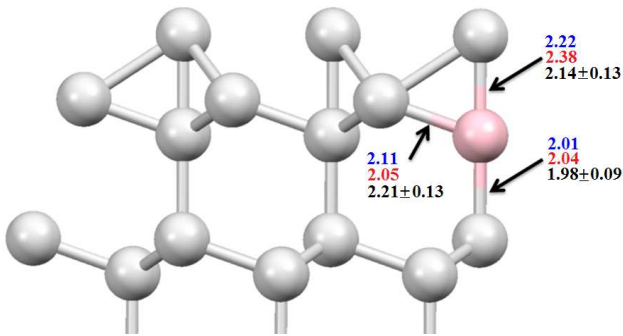}
\caption{Theoretical and experimental bond lengths of for B~atom in
  the L$_{1}^{c}$ position. The numbers stand for lengths in \AA\ as obtained
  by the present work (top), by Andrade \ea\ \cite{Andrade+15}
  (middle), and by Baumg\"{a}rtel \ea\ \cite{Baumg+99}
  (bottom). \label{fig3}}
\end{figure}

The calculations suggest that  L$_{1}^{c}$ is the favourable  configuration. A
detailed view on this configuration together with the bond lengths
obtained from theory and LEED experiments \cite{Baumg+99}
is shown in Figure~\ref{fig3}.  There is a good agreement between
the theoretical and experimental distances and in particular our all-electron
results are always closer to experiment than the pseudopotential results of Andrade
\ea \cite{Andrade+15}.


\subsection{Experimental and theoretical NEXAFS}

\begin{figure}
\includegraphics [width=1.47\linewidth]{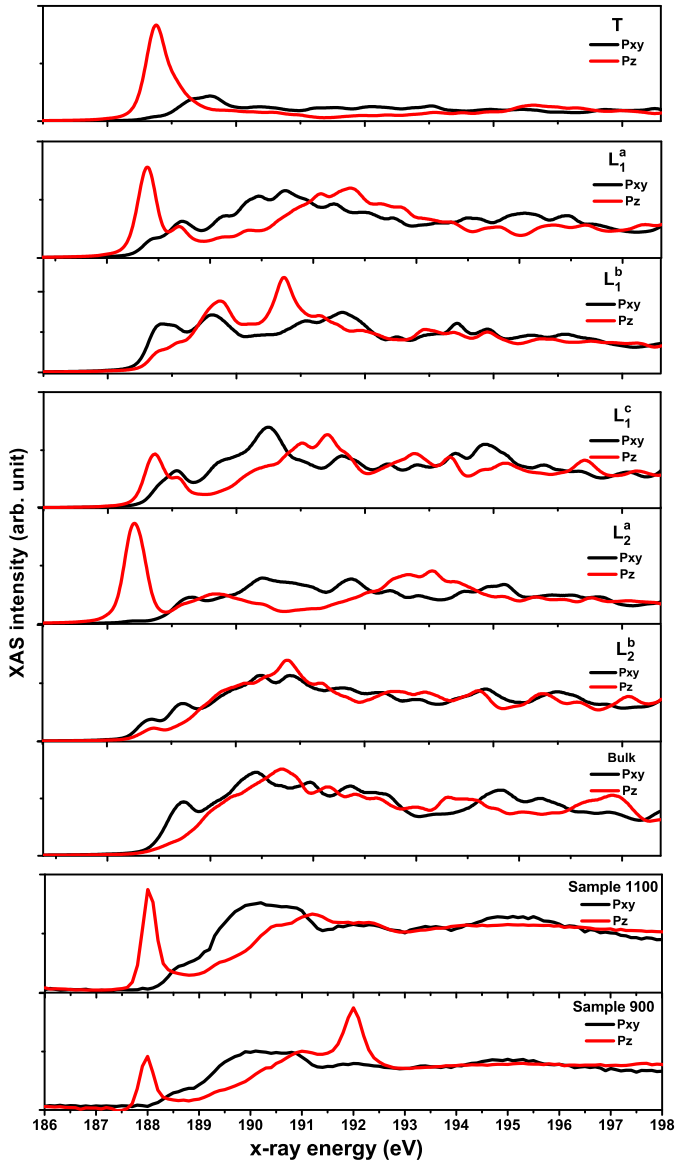}
\caption{Polarized B $K$-edge NEXAFS for B-doped Si(111).  Six upper
  panels show calculated spectra for B~atom in different positions
  denoted in the legend and depicted in Figure~\protect\ref{fig2}.
  Two bottom panels show experimental spectra for the sample 900
  and the sample 1100. \label{fig4}}
\end{figure}

Experimental B $K$-edge spectra for the sample 900 and the sample 1100 of B-doped
Si(111) are shown in the bottom panels of
Figure~\ref{fig4}.  Theoretical spectra obtained for
different positions of B~atoms at or below the reconstructed Si(111)
surface (cf.\ Figure~\ref{fig2}) are shown in the upper panels of
Figure~\ref{fig4}.  Lines identified in the legend as Pxy
stand for spectra with the polarization vector parallel to the surface
(normal incidence), lines identified as Pz stand for spectra with the
polarization vector perpendicular to the surface (grazing
incidence). One can see that the differences between theoretical
spectra for different structural models are large.

When comparing the theory with experiment, one can see that the
L$_{1}^{c}$ model is by far superior to other structural models, both
for the sample 900 and the sample 1100.  However, one should also consider
that the experimental spectra exhibit significant differences between
the samples 900 and 1100.
To get a more complete picture, we performed best-fitting of the
  experimental spectra assuming that 
  the B atoms can be located in various positions simultaneously (see
  Figure~\ref{fig2}).  We employed a fitting procedure which uses
  several criteria for assessing the similarity between the curves, as
  implemented in the {\sc MsSpec} package \cite{SNG+11,msspec-code}. 
The mutual alignment of the spectra originated from different
  sites was performed considering the calculated shifts of the
  energies of the B~1$s$ levels as shown in Tab.~\ref{tab-shifts}.
 These shifts were calculated using the final state rule.

\begin{table}
\caption{Differences between the B~1$s$ core level energies
  for B atoms at different positions as obtained by means of the
  final state rule.  Positive value means that the
  respective 1$s$ electron is bound more strongly than at the
  L$_{1}^{c}$ site. \label{tab-shifts}}
\begin{ruledtabular}
\begin{tabular}{cd}
  \multicolumn{1}{c}{position}
  & \multicolumn{1}{c}{B 1$s$ level shift (eV)} \\ 
\hline 
L$_{1}^{c}$  & 0.00  \\ 
L$_{2}^{a}$  & -0.08  \\ 
L$_{2}^{b}$  & 0.09 \\ 
L$_{1}^{a}$  & -0.01  \\ 
L$_{1}^{b}$  & 0.39  \\ 
T           & 0.07  \\ 
bulk        & 0.31  \\
\end{tabular}
\end{ruledtabular}
\end {table}

\begin{figure}
\includegraphics[width=1.35\linewidth]{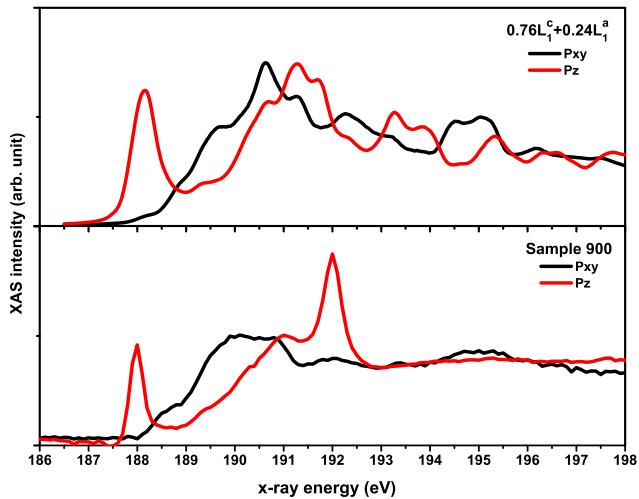}
\caption{Experimental B $K$-edge NEXAFS of the sample 900 compared to
  theoretical NEXAFS for 76~\% of B~atoms in L$_{1}^{c}$ positions and 24~\% of B
  atoms in L$_{1}^{a}$ positions. \label{fig5}}
\end{figure}

A good (though not perfect) fit for the sample 900 is obtained if we
assume that 24~\% of B~atoms are in L$_{1}^{a}$ positions and the rest
in the L$_{1}^{c}$ positions (Figure~\ref{fig5}).  Concerning the
sample 1100, a good is fit obtained by increasing the ratio of
B~atoms in the L$_{1}^{a}$ positions up to 33~\% as shown in Figure~\ref{fig5}). Considering
positions other than L$_{1}^{c}$ or L$_{1}^{a}$ does not improve the
agreement between theory and experiment.
As a whole, we conclude that the majority of B~atoms
occupies the L$_{1}^{c}$ position but a sizable portion of them is 
sitting also somewhere else, possibly in the  L$_{1}^{a}$ position.

Recently it was suggested that for the L$_{1}^{c}$ geometry there may be also
some larger B-Si distances present if two-electron bound states are
formed in the system \cite{Eom+15}. We checked that considering such
geometry (with the atop Si higher above the B~atom than what is shown
in figure~\ref{fig3}) has no significant effect on the calculated
spectra.

\begin{figure}
\includegraphics[width=2.92\linewidth]{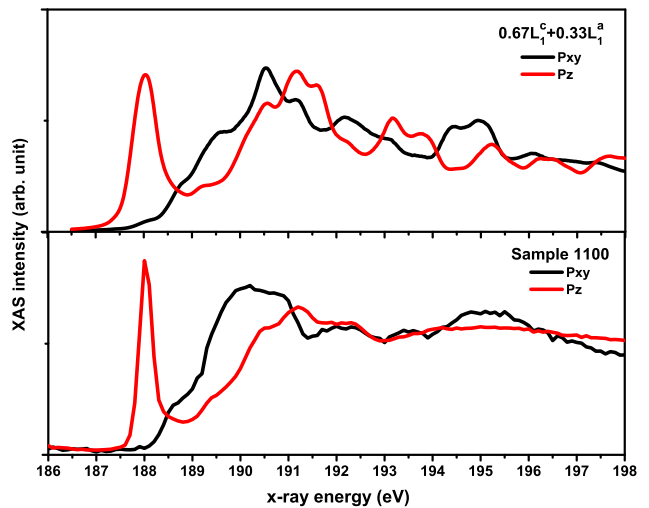}
\caption{Experimental B $K$-edge NEXAFS of the sample 1100 compared to
  theoretical NEXAFS for 67~\% of B~atoms in L$_{1}^{c}$ positions and 33~\% of B
  atoms in L$_{1}^{a}$ positions. \label{fig6}}
\end{figure}


\section{Discussion}

The main goal of the present work was to find the positions of B~atoms at
the Si(111) surface depending on the sample preparation techniques. By
comparing experimental NEXAFS B $K$-edge spectra to spectra calculated
for various model structures we found that the B~atoms are mostly in
the L$_{1}^{c}$ positions. 
However, depending on the preparation method, a significant portion of the B
atoms appears to be in different locations, first of all in the L$_{1}^{a}$ position.

The positions of B atoms as deduced from the NEXAFS experiment
  agree only partially with the total energies calculations.  Most B
  atoms are located in the L$_{1}^{c}$ sites which are also the sites
  with the lowest total energy (Tab.~\ref{tab-totene}).  However, the
  second- and third-lowest energy positions, namely, L$_{2}^{a}$ and
  L$_{2}^{b}$, are not among the sites suggested by the best-fitting
  procedure.  The flashing of the sample used to drive the B atoms
  from the bulk to the surface is apparently a complex non-equilibrium
  procedure and may lead to having the B atoms in metastable positions.

Our NEXAFS-based method is complementary to LEED and STM studies.
This is because with STM studies one can cover always only a small
part of the sample so it is conceivable that in other parts of the
sample the situation may be different than in the part that is
studied.  X-ray absorption spectroscopy and LEED probe much larger
parts of samples so one gets an averaged information concerning the
whole system.  At the same time, unlike LEED, the x-ray absorption
spectroscopy provides a local information because of its chemical
specifity.

Similar to our conclusions, few earlier studies also found that some B
atoms are located in other than L$_{1}^{c}$ positions and that this depends on
the heat treatment \cite{Zotov+96,NMS+00,KOF+11}.  The exact location
of these non-L$_{1}^{c}$ boron atoms is not quite clear and it may further
differ from sample to sample. Our results indicate that for the sample 1100 
 which was subject to flashing at 1100~$^{\circ}$C, some B~atoms
might be at the L$_{1}^{a}$ sites.  However, the agreement between the
experiment and the theory for the sample 1100 is worse than for the sample 900 --- cf.\ figures~\ref{fig5} and~\ref{fig6} --- so our
determination of B~atoms positions for the sample 1100 can be regarded as
tentative.  The calculations predict big differences between NEXAFS
spectra generated for B~atoms in different positions --- see
figure~\ref{fig2}.  Therefore our conclusions concerning
the fact that it is unlikely that a significant portion of B~atoms
would be in the L$_{1}^{b}$, T, L$_{2}^{b}$ and L$_{2}^{a}$ positions are quite robust.
Reckoning all this, it is possbile that some B~atoms in the sample 900
might be associated with surface defects \cite{Spadafora+14} or other
positions not inspected in this work.


\section{Conclusions}

Different preparation conditions of B/Si(111) leads to different
positions of B~atoms at the surface.  Chemically-specific NEXAFS
measurements indicate that most of B~atoms are in the L$_{1}^{c}$ positions, as
it follows also from ab-initio calculations of total energies.
However, for certain preparation conditions and, in particular,
certain modes of heat treatment, a significant portion of B~atoms are
in other positions.  A possible candidate for this other position is
the L$_{1}^{a}$ position --- next to those Si atoms which form the
$(\sqrt{3}\times\sqrt{3})R30^{\circ}$ reconstruction.



\begin{acknowledgments}
 We would like to acknowledge projects CEDAMNF
(CZ.02.1.01/0.0/0.0/15\_003/0000358), LM2015088 and LO1409 of the
Ministry of Education, Youth and Sports (Czech Republic). 
\end{acknowledgments}

\appendix

\section{Core hole effect}

\label{sec-corehole}
\begin{figure}
\includegraphics [width=1.44\linewidth]{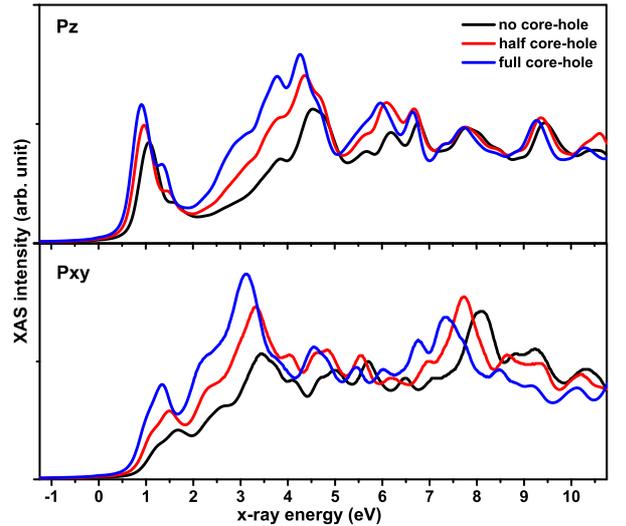}
\caption{Influence of the core hole on the B $K$-edge x-ray absorption
  spectra of B/Si(111). \label{fig7} }
\end{figure}  

We investigate theoretical spectra for the L$_{1}^{c}$ structure for different
ways of including the core hole to see how this influences the
resulting spectra and, based on this, to decide which model is most
suitable for our study.  The calculated B K-edge NEXAFS spectra of
B/Si(111) with no core hole, with half core hole and with a full core
hole are shown in Figure~\ref{fig7}.  One can see that by
varying the strength of the core hole, no new spectral features appear
or disappear for spectra with the polarization vector perpendicular to
the surface (lines denoted as Pz). However for spectra with the
polarization vector parallel to the surface (lines denoted as Pxy), an
extra peak appears near the absorption edge if the strength of the
core hole increases.  Generally, including the core hole leads to an
increase of the intensity of peaks close to the absortion edge.
Besides, a slight shift of peak positions towards lower energies can
be observed.

The full core hole gives the best agreement with experiment for the sample 900 --- compare Figures~\ref{fig4}
and~\ref{fig7}. Therefore we perform all our calculations
using this model.  At the same time, we are aware that our treatment
of the core hole is not perfect and one can expect that including the
core hole in a more elaborate way (beyond the static model) would
probably lead to better results.  For the purpose of distinguishing
between structural models our treatment of the core hole is,
nevertheless, sufficient.

\bibliography{liter-BSi111}

\begin{thebibliography}{33}%
\makeatletter
\providecommand \@ifxundefined [1]{%
 \@ifx{#1\undefined}
}%
\providecommand \@ifnum [1]{%
 \ifnum #1\expandafter \@firstoftwo
 \else \expandafter \@secondoftwo
 \fi
}%
\providecommand \@ifx [1]{%
 \ifx #1\expandafter \@firstoftwo
 \else \expandafter \@secondoftwo
 \fi
}%
\providecommand \natexlab [1]{#1}%
\providecommand \enquote  [1]{``#1''}%
\providecommand \bibnamefont  [1]{#1}%
\providecommand \bibfnamefont [1]{#1}%
\providecommand \citenamefont [1]{#1}%
\providecommand \href@noop [0]{\@secondoftwo}%
\providecommand \href [0]{\begingroup \@sanitize@url \@href}%
\providecommand \@href[1]{\@@startlink{#1}\@@href}%
\providecommand \@@href[1]{\endgroup#1\@@endlink}%
\providecommand \@sanitize@url [0]{\catcode `\\12\catcode `\$12\catcode
  `\&12\catcode `\#12\catcode `\^12\catcode `\_12\catcode `\%12\relax}%
\providecommand \@@startlink[1]{}%
\providecommand \@@endlink[0]{}%
\providecommand \url  [0]{\begingroup\@sanitize@url \@url }%
\providecommand \@url [1]{\endgroup\@href {#1}{\urlprefix }}%
\providecommand \urlprefix  [0]{URL }%
\providecommand \Eprint [0]{\href }%
\providecommand \doibase [0]{http://dx.doi.org/}%
\providecommand \selectlanguage [0]{\@gobble}%
\providecommand \bibinfo  [0]{\@secondoftwo}%
\providecommand \bibfield  [0]{\@secondoftwo}%
\providecommand \translation [1]{[#1]}%
\providecommand \BibitemOpen [0]{}%
\providecommand \bibitemStop [0]{}%
\providecommand \bibitemNoStop [0]{.\EOS\space}%
\providecommand \EOS [0]{\spacefactor3000\relax}%
\providecommand \BibitemShut  [1]{\csname bibitem#1\endcsname}%
\let\auto@bib@innerbib\@empty
\bibitem [{\citenamefont {Lyo}\ \emph {et~al.}(1989)\citenamefont {Lyo},
  \citenamefont {Kaxiras},\ and\ \citenamefont {Avouris}}]{Lyo+89}%
  \BibitemOpen
  \bibfield  {author} {\bibinfo {author} {\bibfnamefont {I.~W.}\ \bibnamefont
  {Lyo}}, \bibinfo {author} {\bibfnamefont {E.}~\bibnamefont {Kaxiras}}, \ and\
  \bibinfo {author} {\bibfnamefont {P.}~\bibnamefont {Avouris}},\ }\href@noop
  {} {\bibfield  {journal} {\bibinfo  {journal} {Phys. Rev. Lett.}\ }\textbf
  {\bibinfo {volume} {63}},\ \bibinfo {pages} {1261} (\bibinfo {year}
  {1989})}\BibitemShut {NoStop}%
\bibitem [{\citenamefont {Headrick}\ \emph {et~al.}(1989)\citenamefont
  {Headrick}, \citenamefont {Robinson}, \citenamefont {Vlieg},\ and\
  \citenamefont {Feldman}}]{Headrick+89}%
  \BibitemOpen
  \bibfield  {author} {\bibinfo {author} {\bibfnamefont {R.~L.}\ \bibnamefont
  {Headrick}}, \bibinfo {author} {\bibfnamefont {I.~K.}\ \bibnamefont
  {Robinson}}, \bibinfo {author} {\bibfnamefont {E.}~\bibnamefont {Vlieg}}, \
  and\ \bibinfo {author} {\bibfnamefont {L.~C.}\ \bibnamefont {Feldman}},\
  }\href@noop {} {\bibfield  {journal} {\bibinfo  {journal} {Phys. Rev. Lett.}\
  }\textbf {\bibinfo {volume} {63}},\ \bibinfo {pages} {1253} (\bibinfo {year}
  {1989})}\BibitemShut {NoStop}%
\bibitem [{\citenamefont {Veiga}\ \emph {et~al.}(2016)\citenamefont {Veiga},
  \citenamefont {Miwa},\ and\ \citenamefont {McLean}}]{Veiga+93}%
  \BibitemOpen
  \bibfield  {author} {\bibinfo {author} {\bibfnamefont {R.~G.~A.}\
  \bibnamefont {Veiga}}, \bibinfo {author} {\bibfnamefont {R.~H.}\ \bibnamefont
  {Miwa}}, \ and\ \bibinfo {author} {\bibfnamefont {A.~B.}\ \bibnamefont
  {McLean}},\ }\href {\doibase 10.1103/PhysRevB.93.115301} {\bibfield
  {journal} {\bibinfo  {journal} {Phys. Rev. B}\ }\textbf {\bibinfo {volume}
  {93}},\ \bibinfo {pages} {115301} (\bibinfo {year} {2016})}\BibitemShut
  {NoStop}%
\bibitem [{\citenamefont {Akimoto}\ \emph {et~al.}(1990)\citenamefont
  {Akimoto}, \citenamefont {Hirsawa}, \citenamefont {Tatsumi}, \citenamefont
  {Hirayama}, \citenamefont {Mizuki},\ and\ \citenamefont
  {Matsui}}]{Akimoto+90}%
  \BibitemOpen
  \bibfield  {author} {\bibinfo {author} {\bibfnamefont {K.}~\bibnamefont
  {Akimoto}}, \bibinfo {author} {\bibfnamefont {I.}~\bibnamefont {Hirsawa}},
  \bibinfo {author} {\bibfnamefont {T.}~\bibnamefont {Tatsumi}}, \bibinfo
  {author} {\bibfnamefont {H.}~\bibnamefont {Hirayama}}, \bibinfo {author}
  {\bibfnamefont {J.}~\bibnamefont {Mizuki}}, \ and\ \bibinfo {author}
  {\bibfnamefont {J.}~\bibnamefont {Matsui}},\ }\href@noop {} {\bibfield
  {journal} {\bibinfo  {journal} {Appl. Physics Lett.}\ }\textbf {\bibinfo
  {volume} {56}},\ \bibinfo {pages} {1225} (\bibinfo {year}
  {1990})}\BibitemShut {NoStop}%
\bibitem [{\citenamefont {Huang}\ \emph {et~al.}(1990)\citenamefont {Huang},
  \citenamefont {Tong}, \citenamefont {Quinn},\ and\ \citenamefont
  {Jona}}]{Huang+90}%
  \BibitemOpen
  \bibfield  {author} {\bibinfo {author} {\bibfnamefont {H.}~\bibnamefont
  {Huang}}, \bibinfo {author} {\bibfnamefont {S.~Y.}\ \bibnamefont {Tong}},
  \bibinfo {author} {\bibfnamefont {J.}~\bibnamefont {Quinn}}, \ and\ \bibinfo
  {author} {\bibfnamefont {F.}~\bibnamefont {Jona}},\ }\href@noop {} {\bibfield
   {journal} {\bibinfo  {journal} {Phys. Rev. B}\ }\textbf {\bibinfo {volume}
  {41}},\ \bibinfo {pages} {3276} (\bibinfo {year} {1990})}\BibitemShut
  {NoStop}%
\bibitem [{\citenamefont {Baumg\"{a}rtel}\ \emph {et~al.}(1999)\citenamefont
  {Baumg\"{a}rtel}, \citenamefont {Paggel}, \citenamefont {Hasselblatt},
  \citenamefont {Horn}, \citenamefont {Fernandez}, \citenamefont {Schaff},
  \citenamefont {Weaver},\ and\ \citenamefont {Bradshaw}}]{Baumg+99}%
  \BibitemOpen
  \bibfield  {author} {\bibinfo {author} {\bibfnamefont {P.}~\bibnamefont
  {Baumg\"{a}rtel}}, \bibinfo {author} {\bibfnamefont {J.~J.}\ \bibnamefont
  {Paggel}}, \bibinfo {author} {\bibfnamefont {M.}~\bibnamefont {Hasselblatt}},
  \bibinfo {author} {\bibfnamefont {K.}~\bibnamefont {Horn}}, \bibinfo {author}
  {\bibfnamefont {V.}~\bibnamefont {Fernandez}}, \bibinfo {author}
  {\bibfnamefont {O.}~\bibnamefont {Schaff}}, \bibinfo {author} {\bibfnamefont
  {J.~H.}\ \bibnamefont {Weaver}}, \ and\ \bibinfo {author} {\bibfnamefont
  {A.~M.}\ \bibnamefont {Bradshaw}},\ }\href@noop {} {\bibfield  {journal}
  {\bibinfo  {journal} {Phys. Rev. B}\ }\textbf {\bibinfo {volume} {59}},\
  \bibinfo {pages} {13014} (\bibinfo {year} {1999})}\BibitemShut {NoStop}%
\bibitem [{\citenamefont {Bedrossian}\ \emph {et~al.}(1989)\citenamefont
  {Bedrossian}, \citenamefont {Meade}, \citenamefont {Mortensen}, \citenamefont
  {Chen}, \citenamefont {Golovchenko},\ and\ \citenamefont
  {Vanderbilt}}]{BMM+89}%
  \BibitemOpen
  \bibfield  {author} {\bibinfo {author} {\bibfnamefont {P.}~\bibnamefont
  {Bedrossian}}, \bibinfo {author} {\bibfnamefont {R.~D.}\ \bibnamefont
  {Meade}}, \bibinfo {author} {\bibfnamefont {K.}~\bibnamefont {Mortensen}},
  \bibinfo {author} {\bibfnamefont {D.~M.}\ \bibnamefont {Chen}}, \bibinfo
  {author} {\bibfnamefont {J.~A.}\ \bibnamefont {Golovchenko}}, \ and\ \bibinfo
  {author} {\bibfnamefont {D.}~\bibnamefont {Vanderbilt}},\ }\href@noop {}
  {\bibfield  {journal} {\bibinfo  {journal} {Phys. Rev. Lett.}\ }\textbf
  {\bibinfo {volume} {63}},\ \bibinfo {pages} {1257} (\bibinfo {year}
  {1989})}\BibitemShut {NoStop}%
\bibitem [{\citenamefont {Kaxiras}\ \emph {et~al.}(1990)\citenamefont
  {Kaxiras}, \citenamefont {Pandey}, \citenamefont {Himpsel},\ and\
  \citenamefont {Tromp}}]{KPH+90}%
  \BibitemOpen
  \bibfield  {author} {\bibinfo {author} {\bibfnamefont {E.}~\bibnamefont
  {Kaxiras}}, \bibinfo {author} {\bibfnamefont {K.~C.}\ \bibnamefont {Pandey}},
  \bibinfo {author} {\bibfnamefont {F.~J.}\ \bibnamefont {Himpsel}}, \ and\
  \bibinfo {author} {\bibfnamefont {R.~M.}\ \bibnamefont {Tromp}},\ }\href@noop
  {} {\bibfield  {journal} {\bibinfo  {journal} {Phys. Rev. B}\ }\textbf
  {\bibinfo {volume} {41}},\ \bibinfo {pages} {1262} (\bibinfo {year}
  {1990})}\BibitemShut {NoStop}%
\bibitem [{\citenamefont {Chang}\ and\ \citenamefont {Stott}(1997)}]{Chang+97}%
  \BibitemOpen
  \bibfield  {author} {\bibinfo {author} {\bibfnamefont {J.}~\bibnamefont
  {Chang}}\ and\ \bibinfo {author} {\bibfnamefont {M.~J.}\ \bibnamefont
  {Stott}},\ }\href@noop {} {\bibfield  {journal} {\bibinfo  {journal} {phys.
  stat. sol. (b)}\ }\textbf {\bibinfo {volume} {200}},\ \bibinfo {pages} {481}
  (\bibinfo {year} {1997})}\BibitemShut {NoStop}%
\bibitem [{\citenamefont {Shi}\ \emph {et~al.}(2002)\citenamefont {Shi},
  \citenamefont {Radny},\ and\ \citenamefont {Smith}}]{Shi+02}%
  \BibitemOpen
  \bibfield  {author} {\bibinfo {author} {\bibfnamefont {H.~Q.}\ \bibnamefont
  {Shi}}, \bibinfo {author} {\bibfnamefont {M.~W.}\ \bibnamefont {Radny}}, \
  and\ \bibinfo {author} {\bibfnamefont {P.~V.}\ \bibnamefont {Smith}},\
  }\href@noop {} {\bibfield  {journal} {\bibinfo  {journal} {Phys. Rev. B}\
  }\textbf {\bibinfo {volume} {66}},\ \bibinfo {pages} {085329} (\bibinfo
  {year} {2002})}\BibitemShut {NoStop}%
\bibitem [{\citenamefont {Andrade}\ \emph {et~al.}(2015)\citenamefont
  {Andrade}, \citenamefont {Miwa}, \citenamefont {Drevniok}, \citenamefont
  {Drage},\ and\ \citenamefont {McLean}}]{Andrade+15}%
  \BibitemOpen
  \bibfield  {author} {\bibinfo {author} {\bibfnamefont {D.~P.}\ \bibnamefont
  {Andrade}}, \bibinfo {author} {\bibfnamefont {R.~H.}\ \bibnamefont {Miwa}},
  \bibinfo {author} {\bibfnamefont {B.}~\bibnamefont {Drevniok}}, \bibinfo
  {author} {\bibfnamefont {P.}~\bibnamefont {Drage}}, \ and\ \bibinfo {author}
  {\bibfnamefont {A.~B.}\ \bibnamefont {McLean}},\ }\href@noop {} {\bibfield
  {journal} {\bibinfo  {journal} {J. Phys.: Condens. Matter}\ }\textbf
  {\bibinfo {volume} {27}},\ \bibinfo {pages} {125001} (\bibinfo {year}
  {2015})}\BibitemShut {NoStop}%
\bibitem [{\citenamefont {Shen}\ \emph {et~al.}(1994)\citenamefont {Shen},
  \citenamefont {Wang}, \citenamefont {Lyding},\ and\ \citenamefont
  {Tucker}}]{Shen+94}%
  \BibitemOpen
  \bibfield  {author} {\bibinfo {author} {\bibfnamefont {T.~C.}\ \bibnamefont
  {Shen}}, \bibinfo {author} {\bibfnamefont {C.}~\bibnamefont {Wang}}, \bibinfo
  {author} {\bibfnamefont {J.~W.}\ \bibnamefont {Lyding}}, \ and\ \bibinfo
  {author} {\bibfnamefont {J.~R.}\ \bibnamefont {Tucker}},\ }\href@noop {}
  {\bibfield  {journal} {\bibinfo  {journal} {Phys. Rev. B}\ }\textbf {\bibinfo
  {volume} {50}},\ \bibinfo {pages} {7453} (\bibinfo {year}
  {1994})}\BibitemShut {NoStop}%
\bibitem [{\citenamefont {Zotov}\ \emph {et~al.}(1996)\citenamefont {Zotov},
  \citenamefont {Kulakov}, \citenamefont {Ryzhkov}, \citenamefont {Saranin},
  \citenamefont {Lifshits}, \citenamefont {Bullemer},\ and\ \citenamefont
  {Eisele}}]{Zotov+96}%
  \BibitemOpen
  \bibfield  {author} {\bibinfo {author} {\bibfnamefont {A.~V.}\ \bibnamefont
  {Zotov}}, \bibinfo {author} {\bibfnamefont {M.~A.}\ \bibnamefont {Kulakov}},
  \bibinfo {author} {\bibfnamefont {S.~V.}\ \bibnamefont {Ryzhkov}}, \bibinfo
  {author} {\bibfnamefont {A.~A.}\ \bibnamefont {Saranin}}, \bibinfo {author}
  {\bibfnamefont {V.~G.}\ \bibnamefont {Lifshits}}, \bibinfo {author}
  {\bibfnamefont {B.}~\bibnamefont {Bullemer}}, \ and\ \bibinfo {author}
  {\bibfnamefont {I.}~\bibnamefont {Eisele}},\ }\href@noop {} {\bibfield
  {journal} {\bibinfo  {journal} {Surf. Sci.}\ }\textbf {\bibinfo {volume}
  {345}},\ \bibinfo {pages} {313} (\bibinfo {year} {1996})}\BibitemShut
  {NoStop}%
\bibitem [{\citenamefont {Stimpel}\ \emph {et~al.}(2000)\citenamefont
  {Stimpel}, \citenamefont {Schulze}, \citenamefont {Hoster}, \citenamefont
  {Eisele},\ and\ \citenamefont {Baumg\"{a}rtner}}]{stimpel}%
  \BibitemOpen
  \bibfield  {author} {\bibinfo {author} {\bibfnamefont {T.}~\bibnamefont
  {Stimpel}}, \bibinfo {author} {\bibfnamefont {J.}~\bibnamefont {Schulze}},
  \bibinfo {author} {\bibfnamefont {H.~E.}\ \bibnamefont {Hoster}}, \bibinfo
  {author} {\bibfnamefont {I.}~\bibnamefont {Eisele}}, \ and\ \bibinfo {author}
  {\bibfnamefont {H.}~\bibnamefont {Baumg\"{a}rtner}},\ }\href@noop {}
  {\bibfield  {journal} {\bibinfo  {journal} {Appl. Surf. Sci.}\ }\textbf
  {\bibinfo {volume} {162}},\ \bibinfo {pages} {384} (\bibinfo {year}
  {2000})}\BibitemShut {NoStop}%
\bibitem [{\citenamefont {Spadafora}\ \emph {et~al.}(2014)\citenamefont
  {Spadafora}, \citenamefont {Berger}, \citenamefont {Mutombo}, \citenamefont
  {Telychko}, \citenamefont {\v{S}̌vec}, \citenamefont {Majzik}, \citenamefont
  {McLean},\ and\ \citenamefont {Jel\'{\i}nek}}]{Spadafora+14}%
  \BibitemOpen
  \bibfield  {author} {\bibinfo {author} {\bibfnamefont {E.~J.}\ \bibnamefont
  {Spadafora}}, \bibinfo {author} {\bibfnamefont {J.}~\bibnamefont {Berger}},
  \bibinfo {author} {\bibfnamefont {P.}~\bibnamefont {Mutombo}}, \bibinfo
  {author} {\bibfnamefont {M.}~\bibnamefont {Telychko}}, \bibinfo {author}
  {\bibfnamefont {M.}~\bibnamefont {\v{S}̌vec}}, \bibinfo {author}
  {\bibfnamefont {Z.}~\bibnamefont {Majzik}}, \bibinfo {author} {\bibfnamefont
  {A.~B.}\ \bibnamefont {McLean}}, \ and\ \bibinfo {author} {\bibfnamefont
  {P.}~\bibnamefont {Jel\'{\i}nek}},\ }\href@noop {} {\bibfield  {journal}
  {\bibinfo  {journal} {J. Phys. Chem. C}\ }\textbf {\bibinfo {volume} {118}},\
  \bibinfo {pages} {15744} (\bibinfo {year} {2014})}\BibitemShut {NoStop}%
\bibitem [{\citenamefont {Nakamura}\ \emph {et~al.}(2000)\citenamefont
  {Nakamura}, \citenamefont {Masuda},\ and\ \citenamefont {Shigeta}}]{NMS+00}%
  \BibitemOpen
  \bibfield  {author} {\bibinfo {author} {\bibfnamefont {K.}~\bibnamefont
  {Nakamura}}, \bibinfo {author} {\bibfnamefont {K.}~\bibnamefont {Masuda}}, \
  and\ \bibinfo {author} {\bibfnamefont {Y.}~\bibnamefont {Shigeta}},\
  }\href@noop {} {\bibfield  {journal} {\bibinfo  {journal} {Surface Science}\
  }\textbf {\bibinfo {volume} {454--456}},\ \bibinfo {pages} {21} (\bibinfo
  {year} {2000})}\BibitemShut {NoStop}%
\bibitem [{\citenamefont {Kr\"{u}gener}\ \emph {et~al.}(2011)\citenamefont
  {Kr\"{u}gener}, \citenamefont {Osten},\ and\ \citenamefont
  {Fissel}}]{KOF+11}%
  \BibitemOpen
  \bibfield  {author} {\bibinfo {author} {\bibfnamefont {J.}~\bibnamefont
  {Kr\"{u}gener}}, \bibinfo {author} {\bibfnamefont {H.~J.}\ \bibnamefont
  {Osten}}, \ and\ \bibinfo {author} {\bibfnamefont {A.}~\bibnamefont
  {Fissel}},\ }\href@noop {} {\bibfield  {journal} {\bibinfo  {journal} {Phys.
  Rev. B}\ }\textbf {\bibinfo {volume} {83}},\ \bibinfo {pages} {205303}
  (\bibinfo {year} {2011})}\BibitemShut {NoStop}%
\bibitem [{\citenamefont {Eom}\ \emph {et~al.}(2015{\natexlab{a}})\citenamefont
  {Eom}, \citenamefont {Moon},\ and\ \citenamefont {Koo}}]{EOM}%
  \BibitemOpen
  \bibfield  {author} {\bibinfo {author} {\bibfnamefont {D.}~\bibnamefont
  {Eom}}, \bibinfo {author} {\bibfnamefont {C.-Y.}\ \bibnamefont {Moon}}, \
  and\ \bibinfo {author} {\bibfnamefont {J.-Y.}\ \bibnamefont {Koo}},\
  }\href@noop {} {\bibfield  {journal} {\bibinfo  {journal} {Nano Letters}\
  }\textbf {\bibinfo {volume} {15}},\ \bibinfo {pages} {398} (\bibinfo {year}
  {2015}{\natexlab{a}})}\BibitemShut {NoStop}%
\bibitem [{\citenamefont {Tang}\ \emph {et~al.}(1992)\citenamefont {Tang},
  \citenamefont {Freeman},\ and\ \citenamefont {Delley}}]{TFD+92}%
  \BibitemOpen
  \bibfield  {author} {\bibinfo {author} {\bibfnamefont {S.}~\bibnamefont
  {Tang}}, \bibinfo {author} {\bibfnamefont {A.~J.}\ \bibnamefont {Freeman}}, \
  and\ \bibinfo {author} {\bibfnamefont {B.}~\bibnamefont {Delley}},\
  }\href@noop {} {\bibfield  {journal} {\bibinfo  {journal} {Phys. Rev. B}\
  }\textbf {\bibinfo {volume} {45}},\ \bibinfo {pages} {1776} (\bibinfo {year}
  {1992})}\BibitemShut {NoStop}%
\bibitem [{\citenamefont {Ramamoorthy}\ \emph {et~al.}(1999)\citenamefont
  {Ramamoorthy}, \citenamefont {Briggs},\ and\ \citenamefont
  {Bernholc}}]{RBB+99}%
  \BibitemOpen
  \bibfield  {author} {\bibinfo {author} {\bibfnamefont {M.}~\bibnamefont
  {Ramamoorthy}}, \bibinfo {author} {\bibfnamefont {E.~L.}\ \bibnamefont
  {Briggs}}, \ and\ \bibinfo {author} {\bibfnamefont {J.}~\bibnamefont
  {Bernholc}},\ }\href {\doibase 10.1103/PhysRevB.59.4813} {\bibfield
  {journal} {\bibinfo  {journal} {Phys. Rev. B}\ }\textbf {\bibinfo {volume}
  {59}},\ \bibinfo {pages} {4813} (\bibinfo {year} {1999})}\BibitemShut
  {NoStop}%
\bibitem [{\citenamefont {Sweetman}\ \emph {et~al.}(2011)\citenamefont
  {Sweetman}, \citenamefont {Jarvis}, \citenamefont {Danza}, \citenamefont
  {Bamidele}, \citenamefont {Gangopadhyay}, \citenamefont {Shaw}, \citenamefont
  {Kantorovich},\ and\ \citenamefont {Moriarty}}]{SJD+11}%
  \BibitemOpen
  \bibfield  {author} {\bibinfo {author} {\bibfnamefont {A.}~\bibnamefont
  {Sweetman}}, \bibinfo {author} {\bibfnamefont {S.}~\bibnamefont {Jarvis}},
  \bibinfo {author} {\bibfnamefont {R.}~\bibnamefont {Danza}}, \bibinfo
  {author} {\bibfnamefont {J.}~\bibnamefont {Bamidele}}, \bibinfo {author}
  {\bibfnamefont {S.}~\bibnamefont {Gangopadhyay}}, \bibinfo {author}
  {\bibfnamefont {G.~A.}\ \bibnamefont {Shaw}}, \bibinfo {author}
  {\bibfnamefont {L.}~\bibnamefont {Kantorovich}}, \ and\ \bibinfo {author}
  {\bibfnamefont {P.}~\bibnamefont {Moriarty}},\ }\href@noop {} {\bibfield
  {journal} {\bibinfo  {journal} {Phys. Rev. Lett.}\ }\textbf {\bibinfo
  {volume} {106}},\ \bibinfo {pages} {136101} (\bibinfo {year}
  {2011})}\BibitemShut {NoStop}%
\bibitem [{\citenamefont {Vasina}(2001)}]{Vasina+01}%
  \BibitemOpen
  \bibfield  {author} {\bibinfo {author} {\bibnamefont {Vasina}},\ }\href@noop
  {} {\bibfield  {journal} {\bibinfo  {journal} {Nucl.\ Inst.\ Methods\ A}\
  }\textbf {\bibinfo {volume} {467--468}},\ \bibinfo {pages} {561} (\bibinfo
  {year} {2001})}\BibitemShut {NoStop}%
\bibitem [{\citenamefont {Blaha}\ \emph {et~al.}(2001)\citenamefont {Blaha},
  \citenamefont {Schwarz}, \citenamefont {Madsen}, \citenamefont {Kvasnicka},\
  and\ \citenamefont {Luitz}}]{Blaha+01}%
  \BibitemOpen
  \bibfield  {author} {\bibinfo {author} {\bibfnamefont {P.}~\bibnamefont
  {Blaha}}, \bibinfo {author} {\bibfnamefont {K.}~\bibnamefont {Schwarz}},
  \bibinfo {author} {\bibfnamefont {G.~K.~H.}\ \bibnamefont {Madsen}}, \bibinfo
  {author} {\bibfnamefont {D.}~\bibnamefont {Kvasnicka}}, \ and\ \bibinfo
  {author} {\bibfnamefont {J.}~\bibnamefont {Luitz}},\ }\href@noop {} {\emph
  {\bibinfo {title} {Wien2k, An Augmented Plane Wave plus Local orbital Program
  for Calculating the Crystal Properties}}},\ \bibinfo {address}
  {\url{http://www.wien2k.at}} (\bibinfo {year} {2001})\BibitemShut {NoStop}%
\bibitem [{\citenamefont {Perdew}\ \emph {et~al.}(1996)\citenamefont {Perdew},
  \citenamefont {Burke},\ and\ \citenamefont {Ernzerhof}}]{Perdew+96}%
  \BibitemOpen
  \bibfield  {author} {\bibinfo {author} {\bibfnamefont {J.~P.}\ \bibnamefont
  {Perdew}}, \bibinfo {author} {\bibfnamefont {K.}~\bibnamefont {Burke}}, \
  and\ \bibinfo {author} {\bibfnamefont {M.}~\bibnamefont {Ernzerhof}},\
  }\href@noop {} {\bibfield  {journal} {\bibinfo  {journal} {Phys. Rev. Lett}\
  }\textbf {\bibinfo {volume} {77}},\ \bibinfo {pages} {3865} (\bibinfo {year}
  {1996})}\BibitemShut {NoStop}%
\bibitem [{\citenamefont {Sun}\ \emph {et~al.}(2015)\citenamefont {Sun},
  \citenamefont {Ruzsinszky},\ and\ \citenamefont {Perdew}}]{Sun+15}%
  \BibitemOpen
  \bibfield  {author} {\bibinfo {author} {\bibfnamefont {J.}~\bibnamefont
  {Sun}}, \bibinfo {author} {\bibfnamefont {A.}~\bibnamefont {Ruzsinszky}}, \
  and\ \bibinfo {author} {\bibfnamefont {J.~P.}\ \bibnamefont {Perdew}},\
  }\href@noop {} {\bibfield  {journal} {\bibinfo  {journal} {Phys. Rev. Lett.}\
  }\textbf {\bibinfo {volume} {115}},\ \bibinfo {pages} {036402} (\bibinfo
  {year} {2015})}\BibitemShut {NoStop}%
\bibitem [{\citenamefont {Tran}\ \emph {et~al.}(2016)\citenamefont {Tran},
  \citenamefont {Stelzl},\ and\ \citenamefont {Blaha}}]{TSB+16}%
  \BibitemOpen
  \bibfield  {author} {\bibinfo {author} {\bibfnamefont {F.}~\bibnamefont
  {Tran}}, \bibinfo {author} {\bibfnamefont {J.}~\bibnamefont {Stelzl}}, \ and\
  \bibinfo {author} {\bibfnamefont {P.}~\bibnamefont {Blaha}},\ }\href@noop {}
  {\bibfield  {journal} {\bibinfo  {journal} {J. Chem. Phys.}\ }\textbf
  {\bibinfo {volume} {144}},\ \bibinfo {pages} {204120} (\bibinfo {year}
  {2016})}\BibitemShut {NoStop}%
\bibitem [{\citenamefont {Monkhorst}\ and\ \citenamefont
  {Pack}(1976)}]{Monkhorst+76}%
  \BibitemOpen
  \bibfield  {author} {\bibinfo {author} {\bibfnamefont {H.~J.}\ \bibnamefont
  {Monkhorst}}\ and\ \bibinfo {author} {\bibfnamefont {J.~D.}\ \bibnamefont
  {Pack}},\ }\href {\doibase 10.1103/PhysRevB.13.5188} {\bibfield  {journal}
  {\bibinfo  {journal} {Phys. Rev. B}\ }\textbf {\bibinfo {volume} {13}},\
  \bibinfo {pages} {5188} (\bibinfo {year} {1976})}\BibitemShut {NoStop}%
\bibitem [{\citenamefont {Mandl}(1992)}]{Mandl+92}%
  \BibitemOpen
  \bibfield  {author} {\bibinfo {author} {\bibfnamefont {F.}~\bibnamefont
  {Mandl}},\ }\href@noop {} {\emph {\bibinfo {title} {Quantum Mechanics}}}\
  (\bibinfo  {publisher} {Wiley},\ \bibinfo {address} {Chichester},\ \bibinfo
  {year} {1992})\BibitemShut {NoStop}%
\bibitem [{\citenamefont {\v{S}ipr}\ and\ \citenamefont {Rocca}(2010)}]{SR+10}%
  \BibitemOpen
  \bibfield  {author} {\bibinfo {author} {\bibfnamefont {O.}~\bibnamefont
  {\v{S}ipr}}\ and\ \bibinfo {author} {\bibfnamefont {F.}~\bibnamefont
  {Rocca}},\ }\href@noop {} {\bibfield  {journal} {\bibinfo  {journal} {J.\
  Synchr.\ Rad.}\ }\textbf {\bibinfo {volume} {17}},\ \bibinfo {pages} {367}
  (\bibinfo {year} {2010})}\BibitemShut {NoStop}%
\bibitem [{\citenamefont {{von Barth}}\ and\ \citenamefont
  {Grossmann}(1982)}]{Barth+82}%
  \BibitemOpen
  \bibfield  {author} {\bibinfo {author} {\bibfnamefont {U.}~\bibnamefont {{von
  Barth}}}\ and\ \bibinfo {author} {\bibfnamefont {G.}~\bibnamefont
  {Grossmann}},\ }\href@noop {} {\bibfield  {journal} {\bibinfo  {journal}
  {Phys. Rev. B}\ }\textbf {\bibinfo {volume} {25}},\ \bibinfo {pages} {5150}
  (\bibinfo {year} {1982})}\BibitemShut {NoStop}%
\bibitem [{\citenamefont {S\'{e}billeau}\ \emph {et~al.}(2011)\citenamefont
  {S\'{e}billeau}, \citenamefont {Natoli}, \citenamefont {Gavaza},
  \citenamefont {Zhao}, \citenamefont {Pieve},\ and\ \citenamefont
  {Hatada}}]{SNG+11}%
  \BibitemOpen
  \bibfield  {author} {\bibinfo {author} {\bibfnamefont {D.}~\bibnamefont
  {S\'{e}billeau}}, \bibinfo {author} {\bibfnamefont {C.}~\bibnamefont
  {Natoli}}, \bibinfo {author} {\bibfnamefont {G.~M.}\ \bibnamefont {Gavaza}},
  \bibinfo {author} {\bibfnamefont {H.}~\bibnamefont {Zhao}}, \bibinfo {author}
  {\bibfnamefont {F.~D.}\ \bibnamefont {Pieve}}, \ and\ \bibinfo {author}
  {\bibfnamefont {K.}~\bibnamefont {Hatada}},\ }\href {\doibase
  http://dx.doi.org/10.1016/j.cpc.2011.07.012} {\bibfield  {journal} {\bibinfo
  {journal} {Comp. Phys. Commun.}\ }\textbf {\bibinfo {volume} {182}},\
  \bibinfo {pages} {2567} (\bibinfo {year} {2011})}\BibitemShut {NoStop}%
\bibitem [{\citenamefont {S\'{e}billeau}(2017)}]{msspec-code}%
  \BibitemOpen
  \bibfield  {author} {\bibinfo {author} {\bibfnamefont {D.}~\bibnamefont
  {S\'{e}billeau}},\ }\href@noop {} {\emph {\bibinfo {title} {The {\sc MsSpec}
  code}}},\ \bibinfo {address}
  {\url{https://ipr.univ-rennes1.fr/msspec?lang=en}} (\bibinfo {year}
  {2017})\BibitemShut {NoStop}%
\bibitem [{\citenamefont {Eom}\ \emph {et~al.}(2015{\natexlab{b}})\citenamefont
  {Eom}, \citenamefont {Moon},\ and\ \citenamefont {Koo}}]{Eom+15}%
  \BibitemOpen
  \bibfield  {author} {\bibinfo {author} {\bibfnamefont {D.}~\bibnamefont
  {Eom}}, \bibinfo {author} {\bibfnamefont {C.-Y.}\ \bibnamefont {Moon}}, \
  and\ \bibinfo {author} {\bibfnamefont {J.-Y.}\ \bibnamefont {Koo}},\ }\href
  {\doibase 10.1021/nl503724x} {\bibfield  {journal} {\bibinfo  {journal} {Nano
  Letters}\ }\textbf {\bibinfo {volume} {15}},\ \bibinfo {pages} {398}
  (\bibinfo {year} {2015}{\natexlab{b}})},\ \bibinfo {note} {pMID:
  25558914}\BibitemShut {NoStop}%
\end{thebibliography}%

\end{document}